\documentclass[amsmath,amssymb, aps, pra,reprint,footinbib,floatfix,superscriptaddress]{revtex4-2}

\usepackage{float}
\usepackage{amsmath,amssymb}
\usepackage[english]{babel}
\usepackage{upgreek}
\usepackage{dsfont}
\usepackage{graphicx}
\usepackage{color}
\usepackage{bbold}
\usepackage{hyperref}
\usepackage[authormarkup=none]{changes}

\begin{document}
\title{Polariton interaction in one-dimensional arrays of atoms coupled to waveguides}

\author{Bj\"{o}rn Schrinski}
\affiliation{Center for Hybrid Quantum Networks (Hy-Q), Niels Bohr Institute, University of Copenhagen, Blegdamsvej 17,
DK-2100 Copenhagen, Denmark}
\author{Anders S. S\o rensen}
\affiliation{Center for Hybrid Quantum Networks (Hy-Q), Niels Bohr Institute, University of Copenhagen, Blegdamsvej 17,
DK-2100 Copenhagen, Denmark}

\begin{abstract}

Photons strongly coupled to material systems constitute a novel system for studying the dynamics of non-equilibrium quantum many-body systems. We give a fully analytical description of the dynamics of  photons coupled to a  one-dimensional array of two-level atoms. The description incorporates input,  scattering inside the medium, and the ejection of  photons from the array. 
We show that inelastic collisions, previously identified in small systems, also occur in infinite systems and elucidate the physics behind this effect. 
The developed theory constitutes an effective field theory for the dynamics, which can be used as a starting point for studies of many-body dynamics.  We discuss different parameter regimes and touch upon possible applications and the limit of many excitations.   

\end{abstract}

\maketitle

\emph{Introduction--}
The interaction between photons  in free space is completely negligible.  As a consequence applications of photon-interaction rely on indirect interaction via non-linear coupling to matter \cite{chang2014quantum,lodahl2015interfacing,roy2017colloquium}. The absorption and re-emission of light in such non-linear media gives rise to a plethora of phenomena like super- \cite{dicke1954coherence,gross1982superradiance,van2013photon} and sub-radiance \cite{van2013photon,asenjo2017exponential,albrecht2019subradiant,zhang2019theory},  electromagnetically induced transparency \cite{fleischhauer2005electromagnetically}, light-matter quantum interfaces \cite{hammerer2010quantum}, single photon transistors \cite{chang2007single},   bound states of light \cite{rupasov1984rigorous,yudson1985dynamics,shen2007strongly,xu2014strongly,bienias2014scattering,mahmoodian2020dynamics,poddubny2020quasiflat,bakkensen2021photonic,iversen2021strongly}, as well as elastic and inelastic scattering \cite{fan2010input,fang2015waveguide,ke2019inelastic}. Understanding such phenomena is of immense technological importance as a crucial ingredient for quantum computation and communication hardware based on photonic systems \cite{chang2014quantum}. At the same time, such systems  with multiple photons constitute an intriguing many body quantum system \cite{hammerer2010quantum,peyronel2012quantum,pichler2015quantum,noh2016quantum}. The exploration of such systems has recently lead to the observation of several effects arising from effective photon-photon interactions  \cite{pritchard2010cooperative,
firstenberg2013attractive,kirchmair2013observation,baur2014single,prasad2020correlating,stiesdal2021controlled}, including photon bound states \cite{liang2018observation}. Even few emitter setups, however, show complex behavior and their description typically requires a large theoretical overhead based on e.g. scattering matrix theory \cite{xu2015input,caneva2015quantum,shi2015multiphoton}, or numerical integration of a wave function Ansatz \cite{fischer2017signatures,trivedi2018few,das2019wave} and effective non-Hermitian Hamiltonians \cite{chang2012cavity,asenjo2017exponential,manzoni2017simulating,albrecht2019subradiant,zhang2019theory}. On the other hand, the cleanest realizations of the  aforementioned phenomena (and many-body dynamics in general) are achieved in the limit of many emitters \cite{chang2012cavity,asenjo2017exponential,albrecht2019subradiant,zhang2019theory,jen2020steady,bakkensen2021photonic}. These systems are also technologically  promising, e.g., for generating photon-photon gates \cite{brod2016passive,schrinski2021phase}. To describe such systems with multiple excitations it is essential to have an efficient description and understanding  of the few-body dynamics, which can then form a  basis for more involved studies of the many-body dynamics.

\begin{figure}
  \centering
  \includegraphics[width=0.35\textwidth]{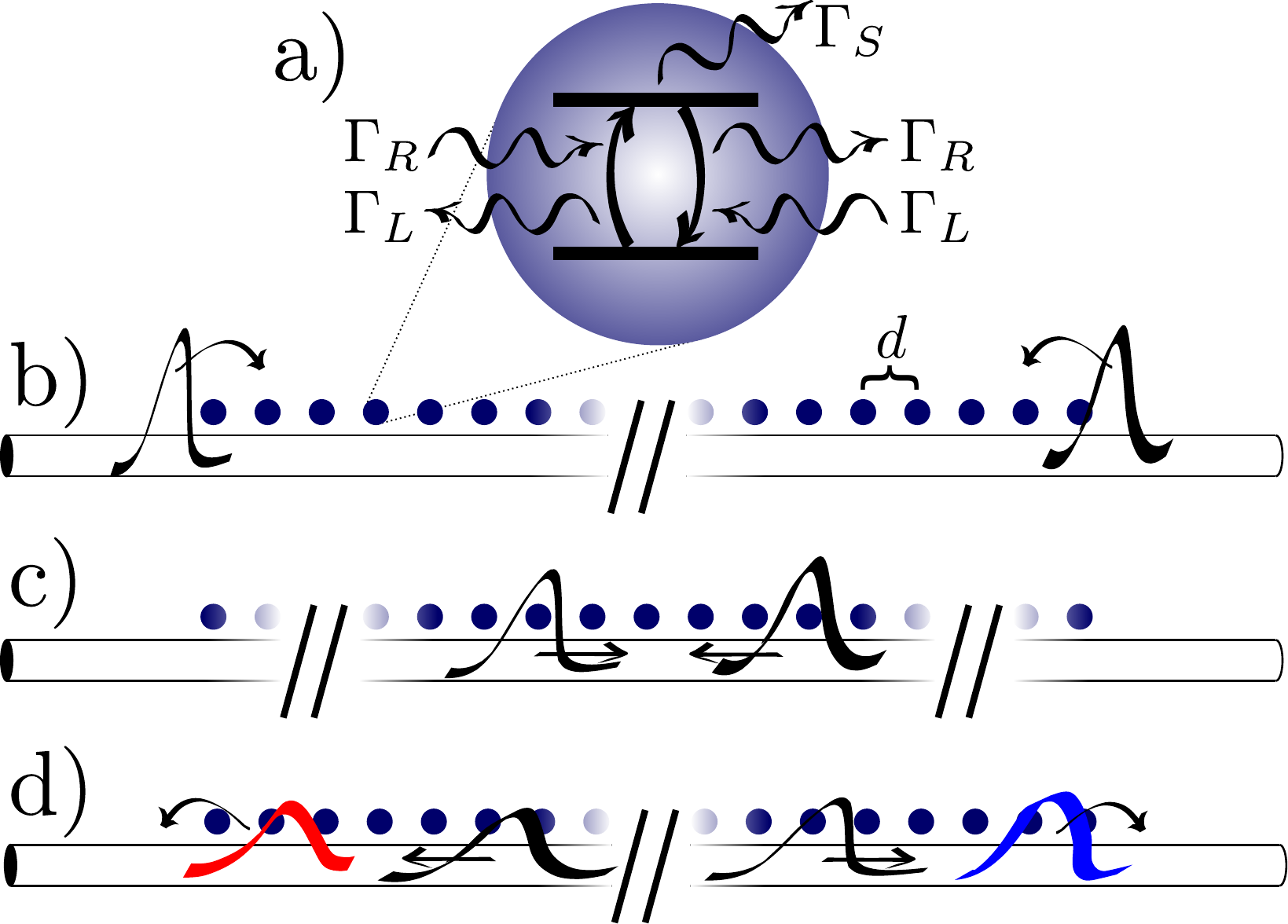}
  \caption{a) Equidistant two level systems with distance $d$ are coupled to a one-dimensional waveguide with decay rates, to the right (left) $\Gamma_R$ ($\Gamma_L$), and to the side $\Gamma_S$. The latter leading to photons being lost entirely from the system. b)$\&$d) Input/output is described by assuming the photons to individually couple to the polariton modes of a semi-infinite array. c) The scattering of polaritons is described inside the chain ignoring any influence from the boundaries. The scattering  may contain inelastic scattering to different photons energies while preserving the total energy.}
\label{fig:Waveguide}
\end{figure}

Here, we  develop a concise fully analytical description of the dynamics of photons coupled to a regular array of multiple two-level emitters, as illustrated in Fig.~\ref{fig:Waveguide} a). This provides an easy to understand picture of the essential dynamics of these systems. Our theory provides a full description of the input and output of the system [Fig.~\ref{fig:Waveguide} (b) and (d)] as well as the propagation and interaction of the excitations [Fig.~\ref{fig:Waveguide} (c)]. Furthermore  we identify an inelastic scattering mechanism where the photons exchange energy, as illustrated in Fig.~\ref{fig:Waveguide} d). Such inelastic collisions have previously been identified for few-emitter systems \cite{ke2019inelastic}. Here we show that such effect also exists in bulk systems and elucidate the physics behind them.    Our results set the stage for explorations of more advanced many-body phenomena in these systems.  We discuss this for the particular example of Tonks-Giradeau gasses \cite{tonks1936complete,girardeau1960relationship}.

\emph{System--}
The system of interest consists of left- and right-propagating waveguide modes with field operaters $\mathcal{E}_L(z)$ and $\mathcal{E}_R(z)$, respectively, with $[\mathcal{E}_{R/L}(z_1),\mathcal{E}^\dagger_{R/L}(z_2)]=c\,\delta(z_1-z_2)$ and $c$ is the group velocity of light in the waveguide. The two level emitter states are described by Pauli operators, e.g.\,$\sigma_\mu^+=|\mathrm{e}\rangle_\mu\langle\mathrm{g}|_\mu$ exciting the $\mu$-th emitter, all with the same distance $d$ to the neighbours.
We consider the typical one-dimensional photon-emitter-interaction Hamiltonian under the rotating wave approximation, $\mathsf{H}_\mathrm{tot}=\mathsf{H}_\mathrm{p}+\mathsf{H}_\mathrm{int}$ with \cite{caneva2015quantum} 
\begin{align}
\mathsf{H}_\mathrm{p}=-i\hbar \int \mathrm{d}z\left[\mathcal{E}_R^\dagger(z)\partial_z\mathcal{E}_R(z)-\mathcal{E}_L^\dagger(z)\partial_z\mathcal{E}_L(z)\right]
\end{align}
and
\begin{align}
\mathsf{H}_\mathrm{int}=\sqrt{\frac{1}{2\pi}}\sum_\mu\left[\mathcal{E}_R(z_\mu) \sqrt{\Gamma_R}+\mathcal{E}_L(z_\mu) \sqrt{\Gamma_L}\right]\sigma_\mu^+ +\mathrm{h.c.}
\end{align}
Here, $\mathsf{H}_\mathrm{tot}$ has been  transformed to the interaction picture absorbing the excitation energy of the emitters. All photon frequencies $\omega$ are thus considered relative to the resonance frequency.  By considering potentially deviating couplings  $\Gamma_L\neq \Gamma_R$ we incorporate an arbitrary level of chirality in the atom chain \cite{lodahl2017chiral}. 

\begin{figure}
  \centering
  \includegraphics[width=0.45\textwidth]{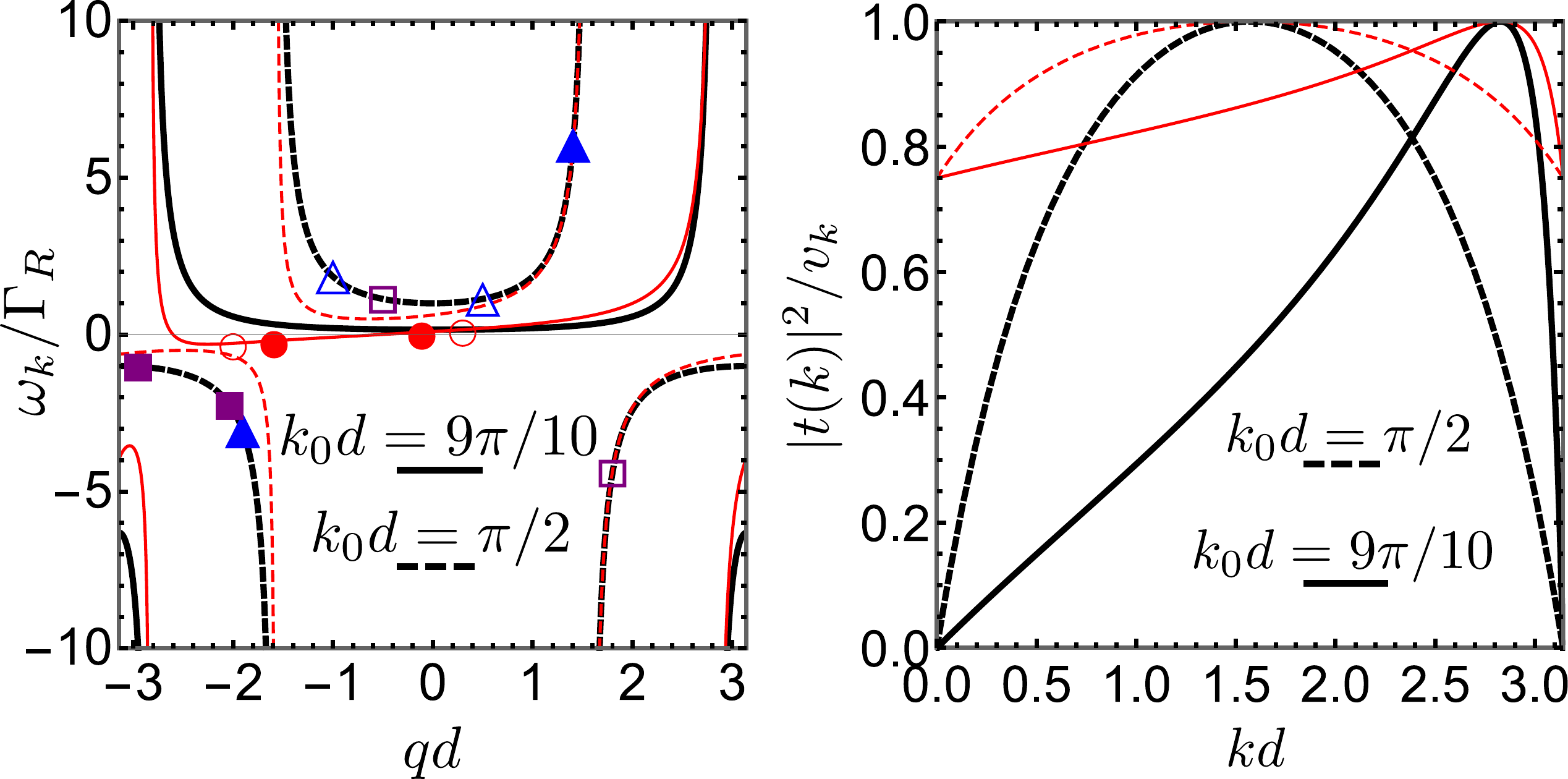}
  \caption{Left: Dispersion relation for different values of $k_0d$ (dashed/solid lines) and chirality, $\Gamma_R/\Gamma_L=1$  (thick black curves) or $4$ (thin red curves). The symbols mark inelastic scattering between incoming (open symbols) into outgoing (filled symbols)  wave numbers. Right: The transmission probability of photons entering the medium or equivalently the probability of polariton modes in the chain being ejected at the boundary. These  depend on $k$, since the frequency is dependent on $k$. For $k$ close to $k_0$ the polaritons are more detuned giving a higher probability to be transmitted. For $k_0d$ close to $0$ or $\pi$ we obtain a linear scaling of the reflection probability leading to a $N^{-3}$ scaling of the longest living subradiant states (see text). For increasing chirality, the transmission $|t(k)|^2/v_k$ increases monotonously due to the suppression of backscattering from each emitter. }
\label{fig:OutputDensity}
\end{figure}

\emph{Polaritons--} A polariton is a quasiparticle emerging from the coupling of a photon to a two level system.
These polaritons can most easily be described after integrating out the photon degree of freedom. Then the dynamics of re-emission and absorption from emitter to emitter is  contained in the effective Hamiltonian \cite{pichler2015quantum,mahmoodian2020dynamics,bakkensen2021photonic}
\begin{align}\label{eq:EffHam}
\mathsf{H}_\mathrm{sys}/\hbar=&-i\sum_{\mu>\nu}e^{ik_0|z_\mu-z_\nu|}\left[\Gamma_R\sigma^+_\mu\sigma^-_\nu+\Gamma_L\sigma^+_\nu\sigma^-_\mu\right]\nonumber\\
&-i\sum_\mu\frac{\Gamma_R+\Gamma_L+\Gamma_S}{2}\sigma^+_\mu\sigma^-_\mu,
\end{align} 
with $k_0$ being the resonance wave number.  The rate $\Gamma_S$ of photons ejected out of the waveguide leads to an overall depletion of the polariton number. While this constant depletion rate can be a considerable limitation for experimental realizations, it does not qualitatively influence the calculations to come apart from introducing a constant loss rate. It will thus be omitted from this point onward. 
For an infinite chain the eigenstates of the system are given by Bloch's theorem  and the effective Hamiltonian \eqref{eq:EffHam} then leads to the dispersion relation 
\begin{align}
\omega_k=-\frac{\Gamma_R}{2}\frac{\cos[(k-k_0)d/2]}{\sin[(k-k_0)d/2]}
+\frac{\Gamma_L}{2}\frac{\cos[(k+k_0)d/2]}{\sin[(k+k_0)d/2]},
\end{align}
plotted in Fig.~\ref{fig:OutputDensity}.

\emph{Input--} We first study the coupling of photons into and out of the atom array as illustrated in Fig.~\ref{fig:Waveguide} b) and d). Using the well-known input-output formalism  \cite{caneva2015quantum} we have
\begin{align}\label{eq:input}
\partial_t \sigma^-_\mu(t)=-i\sqrt{\Gamma_R}\mathcal{E}_\mathrm{in}(z_{1},t)e^{ik_0z_\mu}+\frac{i}{\hbar}[\mathsf{H}_\mathrm{sys},\sigma^-_\mu(t)]
\end{align}
where we without loss of generality chose a right-propagating input field approaching a semi-infinite array.
Taking the spatial Laplace transforms of Eq.~\eqref{eq:input} and the Fourier transform in time we arrive at the input relation for an incoming field $\sigma^-_k=\,t(k)\mathcal{E}_\mathrm{in}(z_1,\omega_k)$ with the transmission amplitude \cite{supp}
\begin{align}
t(k)=-i\sqrt{\Gamma_Rd}[f(k-k_0)+r(k)f(k'-k_0)].
\end{align}
and the Pauli operators in momentum space $\sigma^-_k=\sqrt{d}\sum_\mu e^{ikz_\mu}\sigma^-_\mu$.
Here, we have a factor $f(k-k_0)=1/[1-e^{i(k_0-k)d}]$ and the reflection coefficient is $r(k)=-(e^{-i(k'+k_0)d}-1)/(e^{-i(k+k_0)d}-1)$, where $k'$ is the degenerate wavenumber for which $\omega_k=\omega_{k'}$. 

To obtain a probability density for being coupled into the atom array as a polariton, we  have to correct for the group velocity $v_k=\partial_k\omega_k$ yielding 
\begin{align}
|\sigma^-_k|^2\mathrm{d}k=&\frac{1}{v_k}|t(k)|^2|\mathcal{E}_\mathrm{in}(z_1,\omega_k)|^2 \mathrm{d}\omega_k\nonumber\\
=&\left(1-\frac{v_{k'}}{v_k}|r(k)|^2\right)|\mathcal{E}_\mathrm{in}(z_1,\omega_k)|^2 \mathrm{d}\omega_k.
\end{align}
The transmission  probability $|t(k)|^2/v_k$ becomes smallest for $k\to n\pi$, $n\in\mathbb{Z}$ (i.e.\,the points closest to resonance), where the photon is entirely reflected if $\Gamma_L=\Gamma_R$ (so that $v_{k'}/v_k=1$) expanding the well-known mirror-like behavior for $k_0 d=\pi$\cite{hoi2011demonstration,chang2011multiatomic,chang2012cavity} to general $k_0$, see Fig.~\ref{fig:OutputDensity}. On the other hand, for a completely chiral setup with $\Gamma_L=0$ we get $r(k)=0$ and the photonic wavefunction enters the medium in its entirety. 

\emph{Output--} For the output scenario we start with the output relation  \cite{caneva2015quantum}
\begin{align}\label{eq:output}
\mathcal{E}_\mathrm{out}(z_N,t)=\mathcal{E}_\mathrm{in}(z_N,t)
-i\sqrt{\Gamma_R}\sum_\mu\sigma^-_\mu(t)e^{ik_0(z_N-z_\mu)}
\end{align} 
 with $\mathcal{E}_\mathrm{in}(z_N,t)=0$ for a semi-infinite atom array ending at position $z_N$. Doing a similar transformation as above, we arrive at (see \cite{supp})
\begin{align}
\mathcal{E}_\mathrm{out}(z_N,\omega)
=\frac{1}{|v_{k_\omega}|}e^{i L (k-k_0)}
t_1(k_\omega)\sigma^-_{k_\omega}
\end{align}
analogous to the input scenario, only rescaled by the group velocity. Also, we have an additional phase factor depending on the length $L=d(N-1)$ of the atom array. From here we can calculate the probability for a polariton to leave the array  as a photon
\begin{align}
|E(z_N,\omega)|^2\mathrm{d}\omega
=\left(1-\frac{v_{k'}}{v_k}|r(k)|^2\right)|\sigma^-_{k(\omega)}|^2\mathrm{d}k(\omega).
\end{align} 
The output probability thus exactly matches the input probability in accordance with Helmholtz reciprocity.    

As a direct application, this result allows us to reproduce the previously reported $N^{-3}$ scaling of the subradiant states for $\Gamma_R=\Gamma_L$ \cite{asenjo2017exponential,zhang2019theory}: The lowest energy eigenstates for an array with $N$ atoms are found at  $k=\pi\xi/Nd$ with $\xi\in\mathbb{N}^+$. Wavepackets localized around a certain $k$ will reach the boundaries at a rate $v_k/Nd$. Altogether, this leads to a total ejection rate of
$\gamma=v_k(1-|r(k)|^2)/Nd$ and expanding the transmission $t(k)$ to lowest order in $k$ reproduces the results from Ref.~\cite{zhang2019theory}.

\emph{Polariton scattering--}
The only ingredient missing for a complete picture of the two-photon dynamics is the scattering event shown in Fig.~\ref{fig:Waveguide} c). To describe the scattering of two polaritons we change to center of mass 
and relative coordinates
with absolute momentum $K=(k_1+k_2)/2$ and relative momentum coordinate $q=(k_1-k_2)/2$. The two-excitation momentum basis states can then be expressed as
\begin{align}
|q,K\rangle=\sum_{z_1,z_2}e^{ik_1z_1+ik_2z_2}\sigma^+_{z_1} \sigma^+_{z_2}|0\rangle,
\end{align} 
where $|0\rangle$ is the polariton vacuum state where all atoms are in the ground states $|g\rangle$. 

To solve the scattering problem we apply the effective Hamiltonian \eqref{eq:EffHam} on $|q,K\rangle$ to obtain \cite{bakkensen2021photonic}
\begin{align}\label{eq:EigenEq}
\mathsf{H}_\mathrm{sys}|q,K\rangle=&\hbar\omega_{q,K}|q,K\rangle\nonumber\\
&+\hbar\Gamma_R[i-a(q,k_0-K)]|k_0-K,K\rangle\nonumber\\
&+\hbar\Gamma_L[i-a(q,k_0+K)]|k_0+K,K\rangle,
\end{align}
where the total energy $\omega_{q,K}=\omega_{k_1}+\omega_{k_2}$ and $a(q,p)=\sin q d/[\cos qd-\cos pd]$. 
\begin{figure}
  \includegraphics[width=0.45\textwidth]{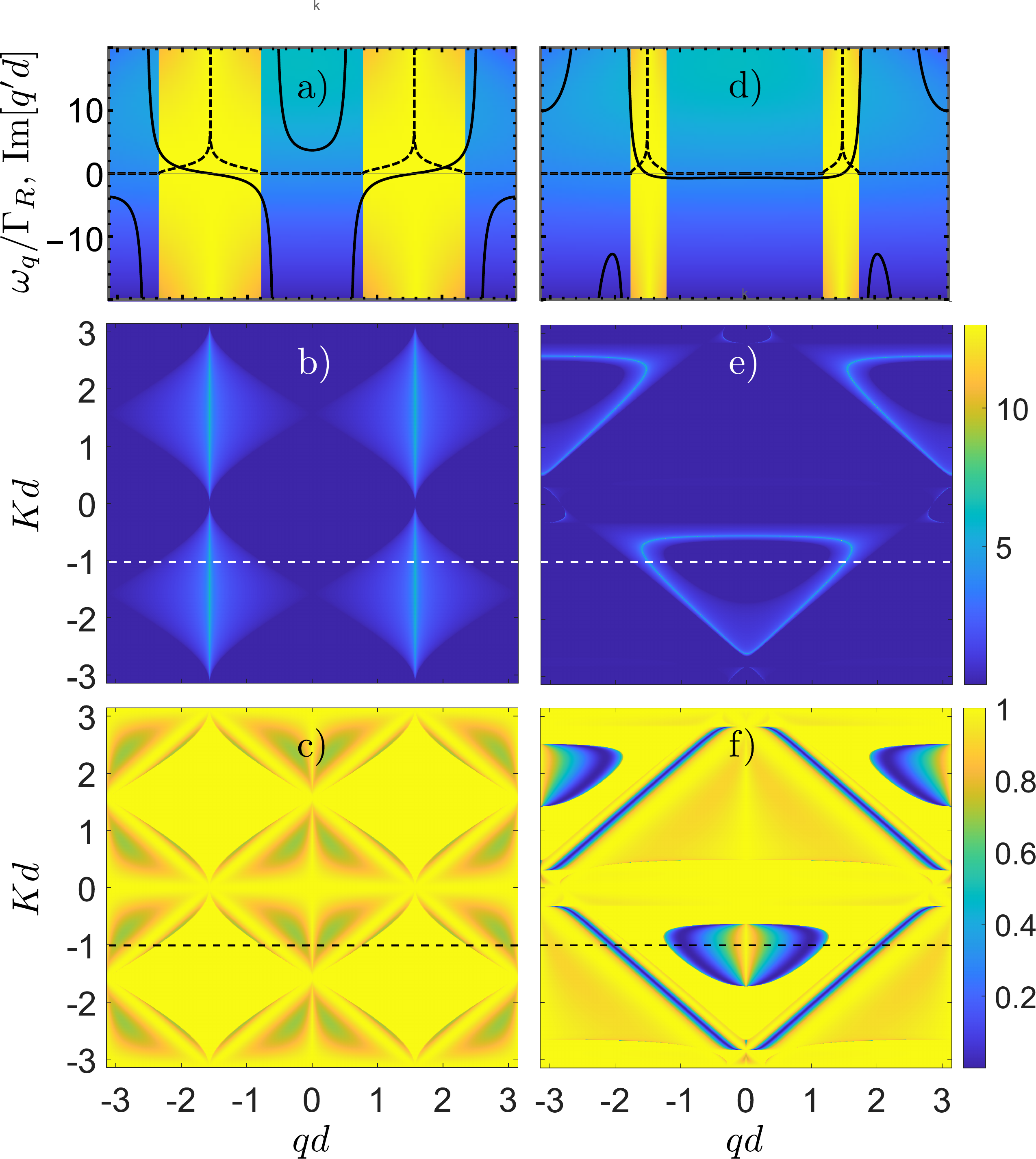}
  \caption{The scattering of two polaritons for different relative and absolute momentum $q$ and $K$ with  a)-c) $k_0d=\pi/2$ and $\Gamma_R/\Gamma_L=1$, d)$-$f) $k_0d=0.9\,\pi$ and $\Gamma_R/\Gamma_L=4$. a)$\&$d): the two-polariton dispersion relation $\omega_{q,K}$ (solid lines) for a fixed $Kd =-1$ marked by dashed lines in b), c), e), and f). For energy regions with four-fold degeneracy (blue region) we have inelastic scattering with $|t_1|^2\leq1$. In d) instead of a minimum at $qd=0$ we have a tiny local maximum in energy which secures a broad region of inelastic scattering. In the yellow regions we require complex $q'$ to solve the eigenequation \eqref{eq:EigenEq} and the imaginary part is plotted as dashed-dotted lines. 
  b)$\&$e): imaginary part of the degenerate $q'$ into which the scattering takes place. Whenever this imaginary part is non-zero we have scattering into meta-stable resonance states with finite extension, while for $\mathrm{Im}\{q'd\}=0$ we have a combination of elastic and inelastic scattering. The maxima of $q'$ in the density plots are actually weakly ascending poles resembling strongly localized resonance states \cite{bakkensen2021photonic}.
  c)$\&$f): The transmission probability $|t_1|^2\leq1$ for elastic scattering into modes with $-q$ as a function of $q$ and $K$.  When  $\mathrm{Im}\{q'd\}\neq0$ the scattering only goes to localized resonance states, so that  $|t_1|^2=1$ is required  to conserve probability \citep{bakkensen2021photonic}, whereas  $|t_1|^2<1$ for $\mathrm{Im}\{q'd\}= 0 $. }
\label{fig:ElasticInelastic}
\end{figure}

To get rid of the second and third line in Eq.~\eqref{eq:EigenEq} and turn it into an eigen-equation we make a scattering ansatz $|\Psi_{q,K}\rangle=|q,K\rangle+t_1|-q,K\rangle+t_2|-q',K\rangle$, where $q'$ is a degenerate momentum number fulfilling $\omega_{q',K}=\omega_{q,K}$ with $|q'|\neq |q|$ (see below). Requiring  $|k_0-K,K\rangle$ and $|k_0+K,K\rangle$ to vanish from $\mathsf{H}_\mathrm{sys} |\Psi_{q,K}\rangle$, then give two equations determining $t_1$ and $t_2$ as a function of $q,K,k_0$ and $\Gamma_R,\Gamma_L$. 
The complete solutions to these equations are given in the supplementary material   \cite{supp}.

In contrast to scattering in free space, the scattering amplitudes depend not only on the relative momentum $q$ but also on their mean momentum $K$. The nature of the scattering depends heavily on the degeneracy of the total energy $\omega_{q,K}$ shown in Fig.~\ref{fig:ElasticInelastic} a) and d). By symmetry the opposite relative momenta $q$ and $-q$ always have the same energy $\omega_{q,K}=\omega_{-q,K}$, but in some regions of phase space [marked with blue in Fig. \ref{fig:ElasticInelastic} a) and d)]  there is  a four-fold degeneracy with real $q$, $-q$, $q'$, and $-q'$ fulfilling $\omega_{q,K}=\omega_{q',K}$. In this case the scattering has an elastic component with a probability $|t_1|^2$ to scatter  into $|-q\rangle$ and an inelastic component scattering into $|-q'\rangle$ with a probability $|t_2|^2v_{q',K}/v_{q,K}$; we show in the supplementary material \cite{supp} that the continuity equation $|t_1|^2+|t_2|^2 v_{q',K}/v_{q,K}=1$ holds. Here $v_{q,K}=\partial_q\omega_{q,K}$ is the relative group velocity and the  probabilities are symmetric under $q\to-q$ or $q\leftrightarrow q'$. 

Physically an inelastic collision means that the photons will be going out  with  different individual  momenta and thus also  different energies. Experimentally such events could thus easily be measured and represent a peculiar form of four wave mixing, which could, e.g., be used for completely frequency conversion of two single photons by picking $t_1\to 0$, e.g.\,the dark blue areas in Fig.~\ref{fig:ElasticInelastic} f).  
In most cases this change of energy results in a sign flip of the energy (relative to resonance $\omega=0$) such that  one of the photons switches to a different  branch of the dispersion relation, see Fig.~\ref{fig:OutputDensity}. In case of a non-chiral setup this is always the case as the two branches are either purely convex or concave, and we cannot have momentum and energy conservation without branch-hopping. For strong chirality, however,  this convexity can be broken and inelastic scattering can take place within the branches, see the red dots in Fig.~\ref{fig:OutputDensity} which are part of the region of inelastic scattering inside the triangle of Fig.~\ref{fig:ElasticInelastic} e).

As opposed to the situation above, for some  combinations of $q,K$ [marked with yellow in Fig.~\ref{fig:ElasticInelastic} a) and d)] the two-polariton dispersion relation  only shows a
two fold degeneracy $\omega_{q,K}=\omega_{-q,K}$. This means that we cannot directly solve the eigenequation \eqref{eq:EigenEq} with \emph{real} relative momenta, and thus inelastic scattering. In this case, however, there exist a different degeneracy $\omega_{q,K}=\omega_{q',K}$, where $q'$ becomes complex, see Fig.~\ref{fig:ElasticInelastic} b)$\&$e).  The wave number $q'$ 
has a positive imaginary part, which leads to a localized, normalizable wave function. Physical such states represent scattering resonances,  where the two polaritons bind together for a finite time \cite{bakkensen2021photonic}. In the regions, where we have scattering resonances, we only have a single outgoing states $|-q\rangle$ and we therefore find $|t_1|^2=1$ as required to conserve probability.

Whether inelastic scattering or scattering into resonance states takes place depends on the setup parameters $k_0,\Gamma_R/\Gamma_L$ and the incoming momenta  $K,q$. In general there is a rather complex interplay between these different effects,  which is visualized in Fig.~\ref{fig:ElasticInelastic} c) \& f).   

\begin{figure}
  \centering
  \includegraphics[width=0.45\textwidth]{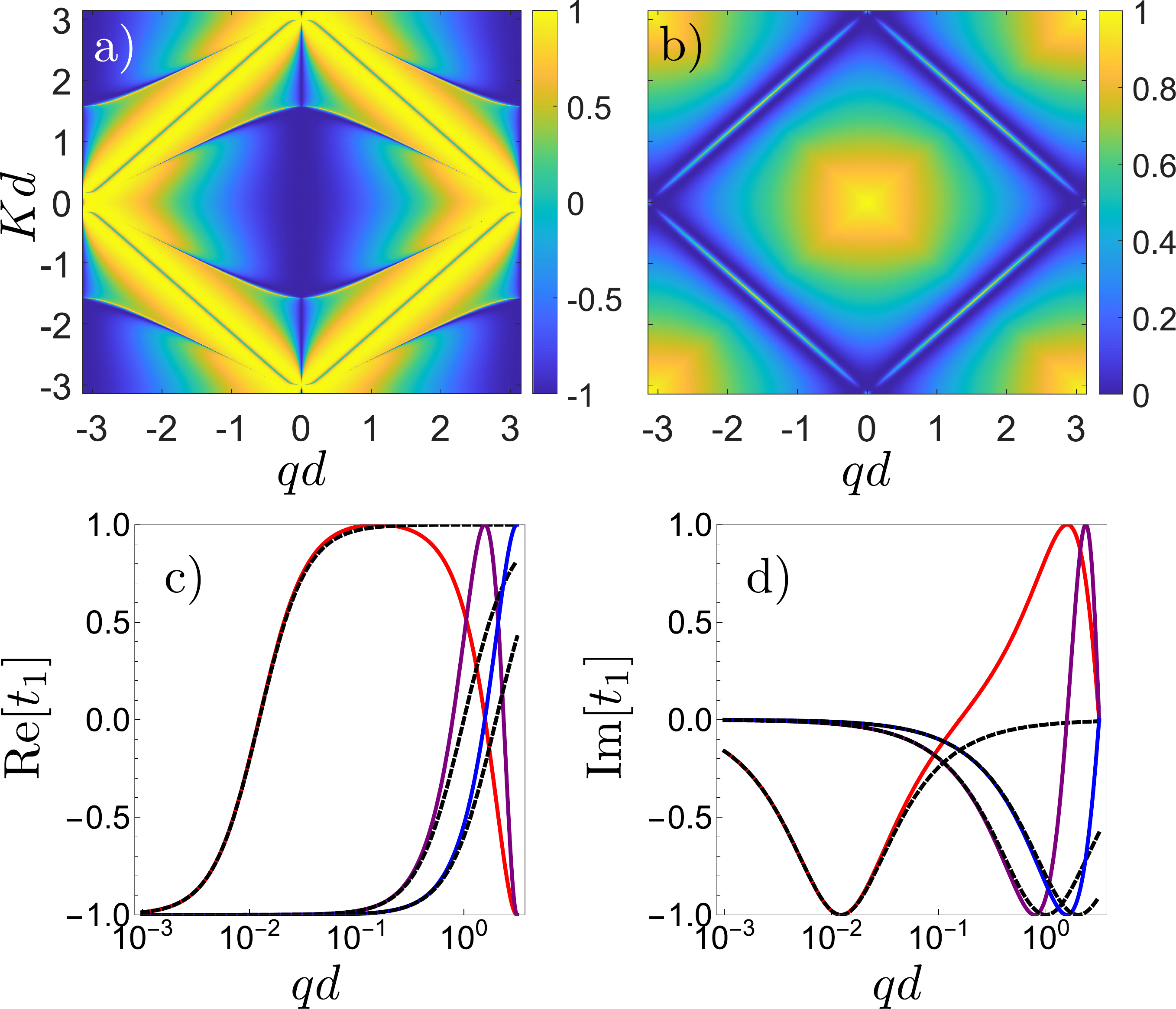}
  \caption{: a) Real part of the transmission coefficient for elastic scattering $t_1$ for $k_0d=0.95\pi,\Gamma_R/\Gamma_L=1$. In the center, we observe a broad region of pure elastic scattering with $t_1=-1$. b) Product of reflection probabilities $|r(k_1)|^2|r(k_2)|^2$ at the boundaries. Near $K,q\approx 0$  we have  $|r(K_1)|^2|r(k_2)|^2\approx1$ so that polaritons are  trapped inside the ensemble facilitating the creation of a Tonks-Girardeu gas. c)$\&$d) Exact real and imaginary part of $t_1$, respectively, in our setup at $K=0$ with $k_0d=0.05\pi$ (red), $k_0d=0.5\pi$ (purple), and $k_0d=\pi$ (blue). The dotted lines are the result of the short range interaction assumed in the Lieb-Liniger model. While this provides the best description for small $k_0$ the broadest region of $t_1=-1$ is found for $k_0d\to\pi$ (although still being restricted to very small $q$).  All plots are for a non-chiral setup $\Gamma_R=\Gamma_L$. }
\label{fig:TonksGirardeuMap}
\end{figure}

\emph{Many particle limit--}
An interesting application of the above results is to use them as a starting point for studying  more complex many-body dynamics in a  long chain of atoms. A particular example of such systems is the Lieb-Lininger model describing    massive particles with short range interactions in one dimension \cite{lieb1963exact}. Here the dynamics is   governed by a scattering length $a$. 
In the limit $a\to 0$ the ground state of this model is a Tonks-Girardeu gas  describing impenetrable particles and fermionization of bosons \cite{tonks1936complete,girardeau1960relationship}.  Below we show how our result predict that a similar behavior can also be produced in our setup for a non-chiral system $\Gamma_R=\Gamma_L=\Gamma_0$. 

To resemble the situation in the Lieb-Liniger model, we are interested in a situation where we only have elastic scattering. The most promising approach is thus  to stay as close to $k\to0$ (corresponding to $K,q\to 0$) as possible, where we are dominated by elastic scattering, see \ref{fig:TonksGirardeuMap} a). By expanding  $\omega_q$ quadratically at the origin we find that the polaritons in this region can be described as massive particles with an effective mass $m_e=2\hbar(1-\cos k_0d)^2/d^2\Gamma_0\sin k_0d$. Furthermore in this region we have near perfect reflection at the boundaries, as shown in \ref{fig:TonksGirardeuMap} b). This ensures that the polaritons will be trapped inside the system enabling the study of long term many-body dynamics. 

To fully match the dynamics of  the Lieb-Liniger model we also need to ensure that the scattering dynamics match the model. For $K=0$ the solution is  especially simple and  only involves scattering into $-q$ with an amplitude 
\begin{align}\label{eq:TransmissionVanishingK}
t_1=-\frac{e^{iq d}-\cos k_0 d}{e^{-iqd}-\cos k_0 d},
\end{align}  
and $t_2=0$ independent of the degree of chirality. 
By expanding the amplitude \eqref{eq:TransmissionVanishingK} for small $k$ we find the  scattering length
\begin{align}\label{eq:scatteringlength}
a=\frac{1}{2}\frac{d}{1-\cos k_0d}
\end{align} 
allowing a complete mapping to the Lieb-Lininger model. 

As opposed to the Lieb-Liniger model, however, the model we use here is fundamentally a lattice model. This difference is explored  in Fig.~\ref{fig:TonksGirardeuMap} c)$\&$d) where we compare  the scattering amplitude with the Lieb-Lininger model. As seen in the figure the resemblance is best for small $k_0d$.  On the other hand,  to maximize the region of phase space with $t_1=-1$ consistent with Tonks-Girardeu-like behavior, it may be desirable to work in a regime with $k_0d\to \pi$. 

In total the mapping to the Lieb-Liniger model presented above indicate that it is indeed possible to  realize a Tonks-Giradeau gas in these system. The analytical theory derived here  provides a firm starting point for a more detailed analysis of this fascinating possibility, but a complete analysis is   beyond the scope of this paper.

\emph{Conclusion--}  
We have described the dynamics of polaritons in atom arrays in the limit of many two-level emitters. 
Our concise analytical model allows for an precise grasp of the rich physics in the system, featuring an interplay of elastic and inelastic scattering as well as scattering into scattering resonances. The results we obtain allow a better understanding of previously obtained results for reflection on resonance \cite{chang2012cavity} and sub-radiance \cite{albrecht2019subradiant,zhang2019theory}.  
At the same time the analytical description of the scattering process paves the way for understanding complex many-body effects. We have outlined a first step towards this by mapping the system to the Lieb-Liniger model and the Tonks-Girardeu gas. This is, however, only one out of the many possibilities. In particular the analytical approach brought forward in this paper may be extended to more involved setups including three-level emitters with deviating chiral coupling \cite{iversen2021strongly} and atom arrays in two or three dimensions. Such extensions may widen the range of many body dynamics, which may be realized with these systems. Furthermore, the theory developed here may also   be applied to enhance photonic  quantum technologies \cite{brod2016passive,schrinski2021phase}.

\emph{Acknowledgements--}  
We are gratefull to Yuxiang Zhang and Bastian Bakkensen for useful discussion.
B.\,S.\,is supported by Deutsche Forschungsgemeinschaft (DFG, German Research Foundation), Grant No. 449674892. We acknowledge the support of Danmarks Grundforskningsfond (DNRF 139, Hy-Q Center for Hybrid
Quantum Networks).

\appendix
\onecolumngrid

\section{Coupling of photons in and out of the atom array}

\subsection{Output}

We start with the easier scenario of a polariton coupling out of the atom array. The generalized input-output-formalism for a chain of emitters (see e.g. \cite{caneva2015quantum}) results in the output field 
\begin{align}
E(z,t)=\mathcal{E}_\mathrm{in}(z,t)-i\sqrt{\Gamma_R}\sum_\mu \sigma^-_\mu(t)e^{ik_0|z-z_\mu|}
\end{align}
with $a_\mu(t)=\langle\sigma^-_\mu\rangle_t$. For now we are only interested in what happens at the boundaries. We therefore for simplicity consider a single excitation in a semi-infinite chain with $z_\mu\in(-\infty,z_N]$. Without the homogeneous solution $\mathcal{E}_\mathrm{in}(z,t)$ (we assume a vacuum state here) the output field right after the last emitter at $z=\epsilon\to0^+$ reads
\begin{align}\label{eq:outputfield1}
E(z_N,t)&=-i\sqrt{\Gamma_R}\sum_{\mu=-\infty}^N \sigma^-_\mu(t)e^{-ik_0z_\mu}\nonumber\\&
=\frac{1}{2\pi i}\sqrt{\Gamma_Rd}\int_{-\pi/d}^{\pi/d} \mathrm{d}k\, \sigma^-_k(t)\sum_{\mu=-\infty}^N e^{i(k-k_0)z_\mu}
=\frac{1}{2\pi i}e^{i L (k-k_0)}\sqrt{\Gamma_Rd}\int_{-\pi/d}^{\pi/d} \mathrm{d}k\, \sigma^-_k(t) f(k-k_0),
\end{align} 
after inserting the inverse Z-transform from first to second line to transform the $\sigma^-_\mu=\sqrt{d}\int_{-\pi/d}^{\pi/d} \mathrm{d}k\, a_k/2\pi$ into momentum space. Here we have introduced the abbreviation $f(k-k_0)=(1-\exp[i(k_0-k)d])^{-1}$ \emph{independent} from the chirality properties of the waveguide. $L=d(N-1)$ is the length of the waveguide and gives an overall $k$-dependent phase that has to be considered for the whole process of photons coupling in and out of the atom array.

For a setup of arbitrary chirality $|k\rangle=\sigma^+_k|\mathrm{vac}\rangle$ is not an eigenstate of the effective (non-hermitian) Hamiltonian 
\begin{align}
\mathsf{H}_\mathrm{eff}=-i\Gamma_R\sum_{\mu>\nu}e^{ik_0|z_\mu-z_\nu|}\sigma^+_\mu\sigma^-_\nu-i\Gamma_L\sum_{\mu<\nu}e^{ik_0|z_\mu-z_\nu|}\sigma^+_\mu\sigma^-_\nu-\frac{\Gamma_R+\Gamma_L}{2}\sum_{\mu}\sigma^+_\mu\sigma^-_\mu
\end{align}
in the Hilbertspace of only the emitters with a single excitation. Applying this effective Hamiltonian on $|k\rangle$ with an half infinite chain of emmiters ($\mu\in(-\infty,z_N]$)
leads to 
\begin{align}\label{eq:SingleTails}
\mathsf{H}_\mathrm{eff}|k\rangle
=\underbrace{\left[-\frac{\Gamma_R}{2}\frac{\cos[(k-k_0)d/2]}{\sin[(k-k_0)d/2]}
+\frac{\Gamma_L}{2}\frac{\cos[(k+k_0)d/2]}{\sin[(k+k_0)d/2]}\right]}_{=\omega_k}|k\rangle
+ie^{iL(k-k_0)}\frac{\Gamma_L}{e^{-i(k+k_0)d}-1}|-k_0\rangle,
\end{align}
with an eigenenergy $\omega_k$. From the dispersion relation we can always find a degenerate $k'$ with $\omega_{k'}=\omega_k$ but this will be different from  $-k$ if a (partially) chiral setup is considered.
From this we can construct eigenstates $|\psi_k\rangle=|k\rangle+r(k)e^{iL(k-k')}|k'\rangle$ of the effective Hamiltonian with
\begin{align}
r(k)=-\frac{e^{-i(k'+k_0)d}-1}{e^{-i(k+k_0)d}-1}=
\begin{cases}
1\quad\mathrm{for}\quad k\rightarrow 0,\pi\\
0\quad\mathrm{for}\quad k\rightarrow k_0.
\end{cases},
\end{align}
with the latter equality if the system is perfectly non-chiral ($\Gamma_R=\Gamma_L$) and thus $k'=-k$. $r(k)$ is the reflection amplitude with the respective probability $|r(k)|^2v_{k'}/v_k$ of polaritons to be reflected at the boundaries, $v_k=\partial_k\omega_k$ being the group velocity.

In general, plane waves with positive $k$ get reflected as plane wave with negative wave numbers $k'$ which also contribute to Eq.~\eqref{eq:output}. 
The electric field can then be written with help of the eigenstates as 
\begin{align}
E(z_N,t)=&
\frac{1}{2\pi i}e^{i L (k-k_0)}\sqrt{\Gamma_Rd}\int_0^{\pi/d}\mathrm{d}k \left[f(k-k_0)a_k(t)+f(k'(\omega_k)-k_0)r(k)\sigma^-_{k'(\omega_k)}(t)\right]\nonumber\\
=&\lim_{\epsilon\to 0^+}\frac{1}{2\pi i}e^{i L (k-k_0)}\sqrt{\Gamma_Rd}\int_{-\infty}^\infty \mathrm{d}\omega_k\frac{\sigma^-_{k(\omega_k)}(t=0)}{\omega-\omega_k+i\epsilon}\left[\frac{f(k(\omega_k)-k_0)}{|v_{k(\omega_k)}|}+\frac{r(k(\omega))f(k'(\omega_k)-k_0)}{|v_{k(\omega_k)}|}\right]\nonumber\\
=&-ie^{i L (k-k_0)}\sqrt{\Gamma_Rd}\left[\frac{f(k(\omega)-k_0)}{|v_{k(\omega)}|}+\frac{r(k(\omega))f(k'(\omega)-k_0)}{|v_{k(\omega)}|}\right]\sigma^-_{k(\omega)}(t=0),
\end{align} 
where we used in the second line Laplace transform into frequency space 
\begin{align}
\sigma^-_k(\omega)=\lim_{\epsilon\to 0}\int_0^\infty \mathrm{d}t'\,e^{i\omega t'-\epsilon t'}\sigma^-_k(t')=\frac{i}{\omega-\omega_k}a_k(t=0),
\end{align}
together with a substitution $\partial_k \omega_k=v_k$, and the third line is achieved via the residue theorem.
In case of $\Gamma_R=\Gamma_L=\Gamma_0$ we have $k'(\omega)=-k(\omega)$ and 
\begin{align}
v_{-k(\omega)}=-v_{k(\omega)}=-\Gamma_0d\sin(k_0 d) \sin(k d)/(\cos(k d) - \cos(k_0 d))^2
\end{align}
which simplifies the infinitesimal energy range of emitted photons to 
\begin{align}
|E(z_N,\omega)|^2\mathrm{d}\omega=\frac{2\sin(k d)\sin(k_0 d)}{1-\cos((k+k_0) d)}|\sigma^-_{k(\omega)}(t=0)|^2\mathrm{d}k(\omega)=(1-|r(k)|^2)|\sigma^-_{k(\omega)}(t=0)|^2\mathrm{d}k(\omega).
\end{align}

Here it makes sense to compare to the $~N^{-3}$ decay rate of subradiant states: Assuming a chain length of $L=Nd$ and the lowest energy states $k=\xi\pi/Nd$ with $\xi\in\mathbb{N}^+$ we obtain the emission rate of waves travelling back and forth from one end of a finite chain to the other as
\begin{align}
\gamma&=\frac{v_k}{L}(1-|r(k)|^2)=\frac{\Gamma_0 d\sin(k_0d)\sin(kd)}{Nd(\cos(kd)-\cos(k_0d))^2}\frac{2\sin(k_0d)\sin(kd)}{(1-\cos((k_0+k)d))}\nonumber\\
&=\frac{2\Gamma_0 k^2d^2}{N}\frac{\sin^2(k_0d)}{(1-\cos(k_0d))^3}+\mathcal{O}(k^3d^3)
=\frac{2\pi^2\Gamma_0\xi^2}{N^3
}\frac{\sin^2(k_0d)}{(1-\cos(k_0d))^3}+\mathcal{O}(k^3d^3),
\end{align}
which is in agreement with results found by diagonalization \cite{zhang2019theory}.

\subsection{Input}

For the input scenario in the perfectly chiral case we can start with a semi-infinite emitter chain with $z_\mu\in[0,\infty)$
\begin{align}
\partial_t \sigma^-_\mu(t)=-i\sqrt{\Gamma_R}\mathcal{E}_\mathrm{in}(z_1,t)e^{ik_0z_\mu}+H_{\mu\nu}\sigma^-_\nu(t)
\end{align}
and directly change to momentum space, $\sigma^-_k(t)=\sqrt{d}\sum_\mu e^{-ikz_\mu}\sigma^-_\mu(t)$, for which we need to know the time evolution of the annihilation operators $\sigma^-_k$. We use the full Lindblad master equation of the reduced system \cite{mahmoodian2020dynamics}
\begin{align}
\mathcal{L}^\dagger \sigma_k^-=&\frac{i}{\hbar}[\mathsf{H}_\mathrm{eff}^\dagger\sigma_k^--\sigma_k^-\mathsf{H}_\mathrm{eff}]+(\Gamma_R+\Gamma_L)\sum_\mu\sigma_\mu^+\sigma_k^-\sigma_\mu^-
\nonumber\\
&+\sum_{\nu>\mu}\left[\left(\Gamma_Re^{ik_0|z_\mu-z_\nu|}+\Gamma_Le^{-ik_0|z_\mu-z_\nu|}\right)\sigma_\nu^+\sigma_k^-\sigma_\mu^-
+\left(\Gamma_Re^{-ik_0|z_\mu-z_\nu|}+\Gamma_Le^{ik_0|z_\mu-z_\nu|}\right)\sigma_\mu^+\sigma_k^-\sigma_\nu^-\right].
\end{align}
Since $\sigma^-_k=\sqrt{d}\sum_{\lambda=0}^\infty e^{-ikz_\lambda}\sigma^-_\lambda$ this results in
\begin{align}
\mathcal{L}^\dagger \sigma^-_k=&-\frac{\Gamma_R+\Gamma_L}{2}
\sigma^-_k-\sqrt{d}\sum_{\nu>\mu}\left[\sigma^-_\nu\Gamma_Le^{ik_0|z_\mu-z_\nu|-ikz_\mu}+\sigma^-_\mu\Gamma_Re^{ik_0|z_\mu-z_\nu|-ikz_\nu}\right]\nonumber\\
=&-i\underbrace{\left(-\frac{\Gamma_R}{2}\frac{\cos[(k-k_0)/2]}{\sin[(k-k_0)/2]}
+\frac{\Gamma_L}{2}\frac{\cos[(k+k_0)/2]}{\sin[(k+k_0)/2]}\right)}_{=\omega_k}\sigma^-_k+\frac{1}{1-e^{-i(k+k_0) d}}\sigma^-_{-k_0}.
\end{align} 
Thus, for $\mathsf{c}_k=\sigma^-_k+r(k)\sigma^-_{k'}$ we have $\partial_t\mathsf{c}_k=-i\omega_k\mathsf{c}_k$. For this operator the input equation becomes
\begin{align}
\partial_tc_k(t)=&-i\sqrt{\Gamma_Rd}\mathcal{E}_\mathrm{in}(z_1,t)\sum_{\mu=0}^\infty e^{ik_0z_\mu}[e^{-ikz_\mu}+r(k)e^{ik'z_\mu}]-i\omega_kc_k(t)\nonumber\\
=&-i\sqrt{\Gamma_Rd}\mathcal{E}_\mathrm{in}(z_1,t)[f(k-k_0)+r(k)f(k'-k_0)]-i\omega_kc_k(t),
\end{align}
and isolation of $c_k(t)$ leads after back and forth transformation between time and frequency (analoguously to the output calculation) to
\begin{align}
c_k(t)=-i\sqrt{\Gamma_Rd}\mathcal{E}_\mathrm{in}(z_1,\omega_k)[f(k-k_0)+r(k)f(k'-k_0)]e^{-i\omega_kt},
\end{align}
The entrance probability for a photon into the chain can be calculated just like in the output case as
\begin{align}
|c_k(t=0)|^2\mathrm{d}k=\Gamma_R|f(k-k_0)+r(k)f(k'-k_0)|^2|\mathcal{E}_\mathrm{in}(z_1,\omega_k)|^2\left|\frac{\partial k}{\partial\omega}\right|\mathrm{d}\omega=\frac{1}{d}F(k,k_0)|\mathcal{E}_\mathrm{in}(z_1,\omega_k)|^2\mathrm{d}\omega,
\end{align}
with $F(k,k_0)=(1-|r(k)|^2)$ if the setup is completely non-chiral. After sufficiently long time (after which the wave package in question is localized far enough away from the boundaries) the $\sigma^-_{k'}(t)$ in $c_k(t)$ do no longer meaningfully contribute and we can map $c_k\to \sigma^-_k$.

\section{Two polariton scattering}

To solve \eqref{eq:EigenEq} we have to find the degenerate $q'$ which fulfill $\omega_{q,K}=\omega_{q',K}$, i.e.
\begin{align}\label{eq:degeneracy}
&\Gamma_L\frac{\sin[(k_0+K)d]}{\cos[qd ]-\cos[(k_0+K)d]}+\Gamma_R\frac{\sin[(k_0-K)d]}{\cos[ qd ]-\cos[(k_0-K)d]}\nonumber\\
=&
\Gamma_L\frac{\sin[(k_0+K)d]}{\cos[ q'd]+\cos[(k_0+K)d]}+\Gamma_R\frac{\sin[(k_0-K)d]}{\cos[  q'd]-\cos[(k_0-K)d]}.
\end{align} 
This equation leads to a quadratical function $c_1\cos^2q'+c_2\cos q'+c_3=0$ with
\begin{align}\label{eq:QuadCoeff}
c_1=&\frac{\sin[(k_0+K)d]}{\cos dk-\cos[(k_0+K)d]}+r\frac{\sin[(k_0-K)d]}{\cos dk-\cos[(k_0-K)d]}\nonumber\\
c_2=&-a[\cos[(k_0+K)d]+\cos[(k_0-K)d]]-\sin[(k_0+K)d]-g\sin[(k_0-K)d]\nonumber\\
c_3=&a\cos[(k_0+K)d]\cos[(k_0-K)d]+\sin[(k_0+K)d]\cos[(k_0-K)d]+g\sin[(k_0-K)d]\cos[(k_0+K)d],
\end{align}
with $r=\Gamma_R/\Gamma_L$. The solution of this quadratic equation can be simplified to
\begin{align}\label{eq:DegenerateMomenta}
&\frac{r\cos^2[(k_0+K)d]\sin[(k_0-K)d]+\cos qd [(r-1)\sin(2Kd)-(r+1)\sin(2k_0d)]+2\cos^2[(k_0-K)d]\sin[(k_0+K)d]}{2(r-1)\sin Kd[\cos qd \cos k_0d-\cos Kd]+2(1+r)\sin k_0d[\cos k_0d-\cos qd \cos Kd]}\nonumber\\
=&\cos q',
\end{align}
or $q'=\pm q$. Inserting these $q'$ into \eqref{eq:EigenEq} to get rid of the second and third line results in the two conditions
\begin{align}
1+i a(q,k_0+K)+t_1[1-ia(q,k_0+K)]+t_2[1-ia(q',k_0+K)]=&0\\
1+i a(q,k_0-K)+t_1[1-ia(q,k_0-K)]+t_2[1-ia(q',k_0-K)]=&0,
\end{align}
which is solved by
\begin{align}\label{eq:TransmissionCoefficients}
t_1=&\frac{a(q,k_0+K)[a(q',k_0-K)-i]+i[a(q,k_0-K)-a(q',k_0-K)]-a(q',k_0+K)[a(q,k_0-K)-i]}{a(q,k_0+K)[a(q',k_0-K)-i]+i[a(q,k_0-K)+a(q',k_0-K)]-a(q',k_0+K)[a(q,k_0-K)+i]}\\
t_2=&\frac{2i[a(q,k_0+K)-a(q,k_0-K)]}{a(q,k_0+K)[a(q',k_0-K)-i]+i[a(q,k_0-K)+a(q',k_0-K)]-a(q',k_0+K)[a(q,k_0-K)+i]}.
\end{align}
Inserting the degenerate momentum \eqref{eq:TransmissionCoefficients} into the transmission coefficient \eqref{eq:EigenEq} leads to a cumbersome analytical expression. The scattering fulfills the continuity equation with the relative group velocities $v_g(q)=\partial \omega(q)/\partial q$. To show this we note that 
\begin{align}\label{eq:continuityextraterm}
|t_1|^2+|t_2|^2 v_g(q')/v_g(q)-1=
\frac{(a(q',k_0+K)-a(q',k_0-K))v_g(q')/v_g(q)-a(q,k_0+K)+a(q,k_0-K)}{f(q,q',k_0,K)}
\end{align}
for every $q$ with degenerate, \emph{real} $q'$. The complicated function $f(q,q',k_0,K)$ is of no further interest since the numerator is proportional to
\begin{align}
&\frac{\sin[(k_0+K)d](\cos[qd]-\cos[(k_0-K)d])}{\cos[qd ]-\cos[(k_0+K)d]}+r\frac{\sin[(k_0-K)d](\cos[qd ]-\cos[(k_0+K)d])}{\cos[ qd]-\cos[(k_0-K)d]}\nonumber\\
-&\frac{\sin[(k_0+K)d](\cos[q'd]-\cos[(k_0-K)d])}{\cos[q'd]-\cos[(k_0+K)d]}-r\frac{\sin[(k_0-K)d](\cos[q'd]-\cos[(k_0+K)d])}{\cos[q'd]-\cos[(k_0-K)d]}
\end{align}
which is very close to the degeneracy condition \eqref{eq:degeneracy}, leads to a quadratic equation with coefficients proportional to \eqref{eq:QuadCoeff} (the factor being $\cos[(k_0+K)d]-\cos[(k_0-K)d]$) and vanishes when Eq.~\eqref{eq:DegenerateMomenta} holds. Thus for degenerate $k'$ the term \eqref{eq:continuityextraterm} always vanishes and the continuity equation is fulfilled at all degrees of chirality $r=\Gamma_R/\Gamma_L$. Whenever $k'$ is complex, the scattering involves meta-stable states with finite extension and $|t_1|=1$.

\end{document}